\documentclass[twocolumn,prl,preprintnumbers,superscriptaddress]{revtex4-1}
\usepackage{amsmath,amssymb,amsfonts,mathrsfs,braket,amsthm,bm,graphicx,booktabs,xcolor,siunitx}
\usepackage[T1]{fontenc}
\usepackage[varg]{txfonts}
\usepackage{pifont}
\usepackage{wasysym}

\usepackage[colorlinks=true,allcolors=blue,breaklinks]{hyperref}

\def\equationautorefname~#1\null{Eq.~(#1)\null}
\allowdisplaybreaks

\begin{document}

\title{Higher-Order Band Topology in Twisted Bilayer Kagome Lattice}

\author{Xiaolin Wan}
\affiliation{Institute for Structure and Function $\&$ Department of Physics, Chongqing University, Chongqing 400044, P. R. China}
\affiliation{Chongqing Key Laboratory for Strongly Coupled Physics, Chongqing 400044, P. R. China}


\author{Junjie Zeng}
\email[]{jjzeng@cqu.edu.cn}
\affiliation{Institute for Structure and Function $\&$ Department of Physics, Chongqing University, Chongqing 400044, P. R. China}
\affiliation{Chongqing Key Laboratory for Strongly Coupled Physics, Chongqing 400044, P. R. China}

\author{Ruixiang Zhu}
\affiliation{Institute for Structure and Function $\&$ Department of Physics, Chongqing University, Chongqing 400044, P. R. China}
\affiliation{Chongqing Key Laboratory for Strongly Coupled Physics, Chongqing 400044, P. R. China}

\author{Dong-Hui Xu}
\email[]{donghuixu@cqu.edu.cn}
\affiliation{Institute for Structure and Function $\&$ Department of Physics, Chongqing University, Chongqing 400044, P. R. China}
\affiliation{Chongqing Key Laboratory for Strongly Coupled Physics, Chongqing 400044, P. R. China}

\author{Baobing Zheng}
\email{scu$_$zheng@163.com}
\affiliation{College of Physics and Optoelectronic Technology, Baoji University of Arts and Sciences, Baoji 721016, P. R. China}

\author{Rui Wang}
\affiliation{Institute for Structure and Function $\&$ Department of Physics, Chongqing University, Chongqing 400044, P. R. China}
\affiliation{Chongqing Key Laboratory for Strongly Coupled Physics, Chongqing 400044, P. R. China}
\affiliation{Center of Quantum Materials and Devices, Chongqing University, Chongqing 400044, P. R. China}

\begin{abstract}
Topologically protected corner states serve as a key indicator for two-dimensional higher-order topological insulators, yet they have not been experimentally identified in realistic materials. Here, by utilizing the effective tight-binding model and symmetry arguments, we establish a connection between higher-order topological insulators and twisted bilayer kagome lattices. We find that the topologically nontrivial bulk band gap arises in the twisted bilayer kagome lattice system due to twist-induced intervalley scattering, leading to the emergence of higher-order topological insulators with a range of commensurate twist angles, and the higher-order band topology is verified by the second Stiefel-Whitney number and fractionally quantized corner charges. Moreover, we investigate the influence of disorder and charge density wave order on the stability of higher-order topological insulator phases. The results show that the corner states of twisted bilayer kagome lattice systems are robust with respect to disorder and charge density wave. Our work not only provides a feasible approach to realize the readily controllable higher-order topological insulator phases by employing a simple twist technique, but also demonstrates that the twisted bilayer kagome lattice systems exhibit the robustness of higher-order band topology, making it feasible to check above prediction in experiments.
\end{abstract}



\maketitle
\textit{Introduction}.---The field of condensed-matter physics has been revolutionized by the discovery of unexpected electronic behaviors in magic-angle twisted bilayer graphene (TBG)~\cite{Cao2018,Cao2018s}. This breakthrough has given rise to a new area of research known as twistronics, which explores the profound effects of manipulating the relative orientation between two single layers of two-dimensional~(2D) materials. By rotating one layer with respect to the other, complex moir\'e patterns emerge at the atomic scale in TBG. These patterns create a tunable experimental platform exhibiting remarkable emergent quantum phenomena, including low-energy van Hove singularities~\cite{Li2010,PhysRevLett.109.126801}, ferromagnetism~\cite{twist-ferromagnetic,Ferro2020}, the quantum Hall effect~\cite{PhysRevLett.107.216602,PhysRevLett.108.076601}, strongly correlated Mott-like insulators~\cite{Cao2018}, and unconventional superconductivity \cite{Cao2018s,Po2018,Hc2018xx,Yankowitz2019}. Beyond TBG, various twisted moir\'e superlattices have been proposed and found to accommodate intriguing physics~\cite{zhang2017interlayer,Regan2020,Wang2020,Tang2020,PhysRevLett.121.026402,PhysRevLett.122.086402,PhysRevB.99.125424,McGilly2020,Claassen2022,PhysRevLett.132.086302}.

Another aspect of twisted systems is their nontrivial band topology. Moir\'e superlattices can induce the occurrence of topological Chern bands with tunable Chern numbers, resulting in the intrinsic quantum anomalous Hall effect~\cite{PhysRevB.99.075127,doi:10.1126/science.aay5533,PhysRevLett.124.166601,chernflatband2022,VCQAHEprx2024,Tflatband2024}. Moreover, the interplay between flatband topology and strong electron correlations can lead to fractional Chern insulators \cite{FCI2020,FCI2021}. Recently, the concept of band topology has been generalized to higher-order topological phases of matter~\cite{Benalcazar2017,Song2017,Benalcazar2017a,Langbehn2017,Schindler2018,Peterson2018,Ezawa2018,Wang2023xxx,zjj2022}. In a higher-order topological insulator (HOTI), the dimensionality of nontrivial boundary states is more than one dimension below that of the bulk. For instance, in a $d$-dimensional topological insulator of order $n$, the gapless boundary states manifest at boundaries with dimensions ($d-n$), where $n$ is greater than $1$. Numerous higher-order topological states have been theoretically predicted in various systems, including TBG systems~\cite{Schindler2018a,PhysRevLett.123.256402,PhysRevB.98.045125,PhysRevB.98.245102,PhysRevLett.123.186401,Qian2021,Zhan2024,Li2023,Guo2022,Ren2020,Lee2020,Liu2019,Zeng2021,Qian2022,PhysRevLett.123.196401,PhysRevLett.124.036803,Wang2022,Park2019,PhysRevLett.126.066401,THOTI}. While experimental realizations of certain predicted states have been achieved in artificial systems such as acoustic metamaterial \cite{Xue2019,Ni2019,Zhang2019}, photonic crystals \cite{PhysRevLett.122.233902,PhysRevLett.122.233903}, electrical circuit \cite{Imhof2018,PhysRevB.99.020304,Song2020}, a crucial challenge persists: the definitive experimental observation of corner states in 2D HOTIs constructed from naturally occurring solid-state materials remains elusive.

The kagome lattice, a geometrically rich network of interconnected triangles and hexagons, has emerged as a versatile platform for investigating exotic quantum phenomena~\cite{RevModPhys.89.025003}.
Recently, kagome materials such as the A-V-Sb family (A =K, Rb, and Cs) \cite{Wilson2024,Lin2021,PhysRevLett.127.217601,Nie2022,FENG20211384,Tan2021,Yu2021,Kim2023} and FeSn~\cite{Kang2019,Xie2021} have garnered significant attention due to the intricate interplay of nontrivial band topology, charge density waves (CDW), and superconductivity~\cite{Yin2022,Neupert2022,Wang2023}.
These developments in kagome materials inspire us to investigate higher-order topology in twisted kagome structures combining the unique properties of kagome lattices with twisted physics. Crucially, investigating HOTI phases within 2D twisted kagome moir\'e superlattices could provide a solution to the longstanding challenge of experimentally identifying corner states in real materials. The inherent tunability offered by the twist approach presents a promising pathway for the controllable design and realization of exotic higher-order topological phases.

In this Letter, we employ a combination of effective model calculations and symmetry-based arguments to systematically investigate higher-order band topology in a twisted bilayer kagome lattice (TBKL) across a range of twist angles. We utilize a tight-binding model to elucidate the relationship between band gaps and twist angles, attributing this dependence to twist-induced intervalley scattering. To rigorously establish the presence of higher-order band topology in TBKL systems, we calculated bulk topological indices, including the second Stiefel-Whitney number and fractionally quantized corner charge, as well as conduct a detailed analysis of edge and corner states in nanosheets. Furthermore, we demonstrate the robustness of the higher-order band topology against disorder and CDW, thereby leading to an exciting prospect for its experimental verification.

\begin{figure}
    \centering
    \includegraphics[scale=0.33]{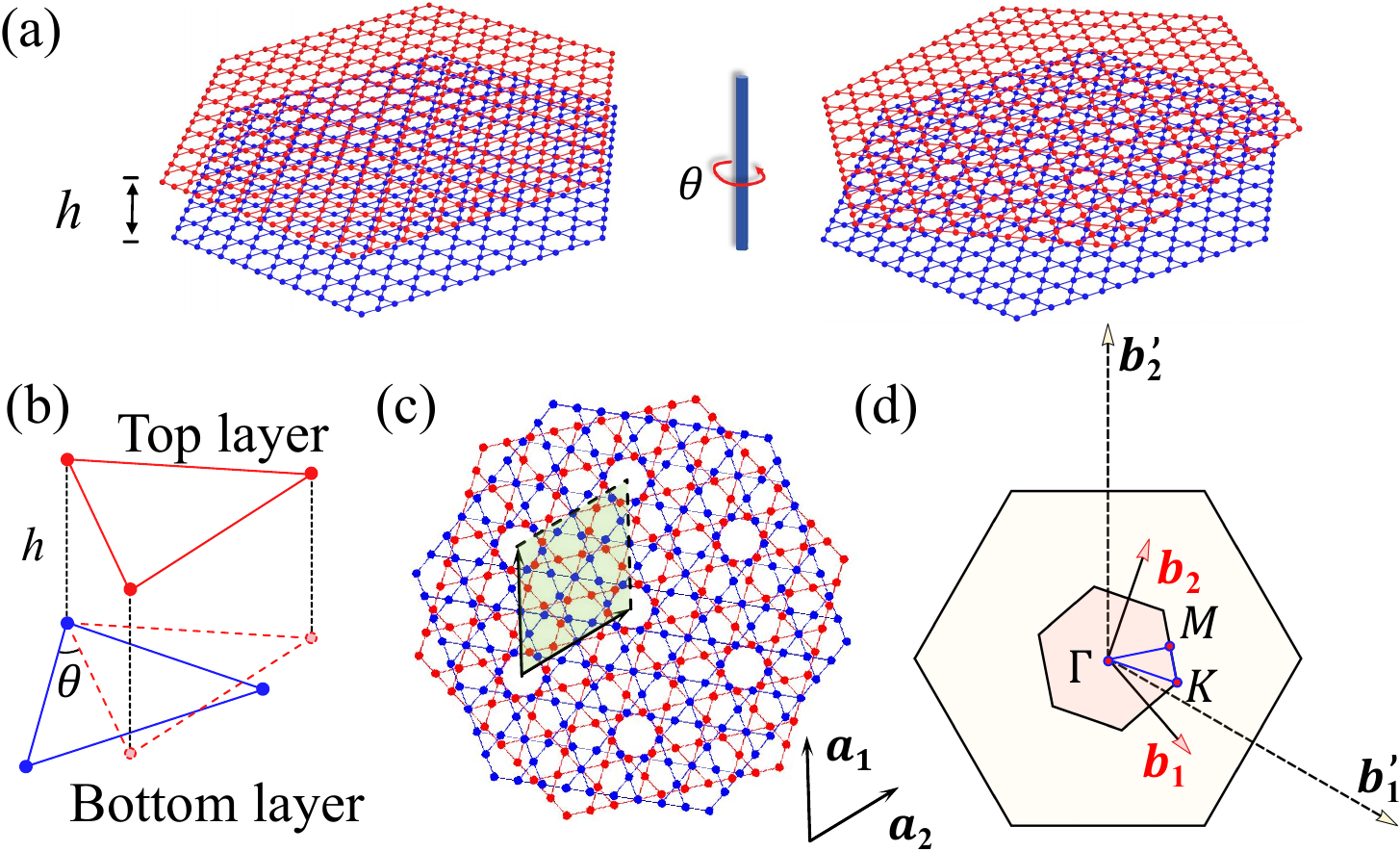}
    \caption{(a) On the left is a schematic of an AA-stacking bilayer kagome lattice with interlayer distance $h$, and on the right is schematic of the corresponding TBKL, in which the top layer is rotated around a certain vertical direction but the bottom layer keeps immobile. (b) Schematic of rotation angle $\theta$. (c) Top view of the TBKL when the twist angle $\theta$ is \ang{21.78}, in which the unit cell and the lattice vectors are marked. (d) The reduced hexagonal Brillouin zone (BZ) of the TBKL system~(in light red), and the original Brillouin zone of the monolayer kagome lattice (in light yellow).
    \label{fig1}}
\end{figure}

\begin{figure*}[t]
    \centering
    \includegraphics[scale=0.092]{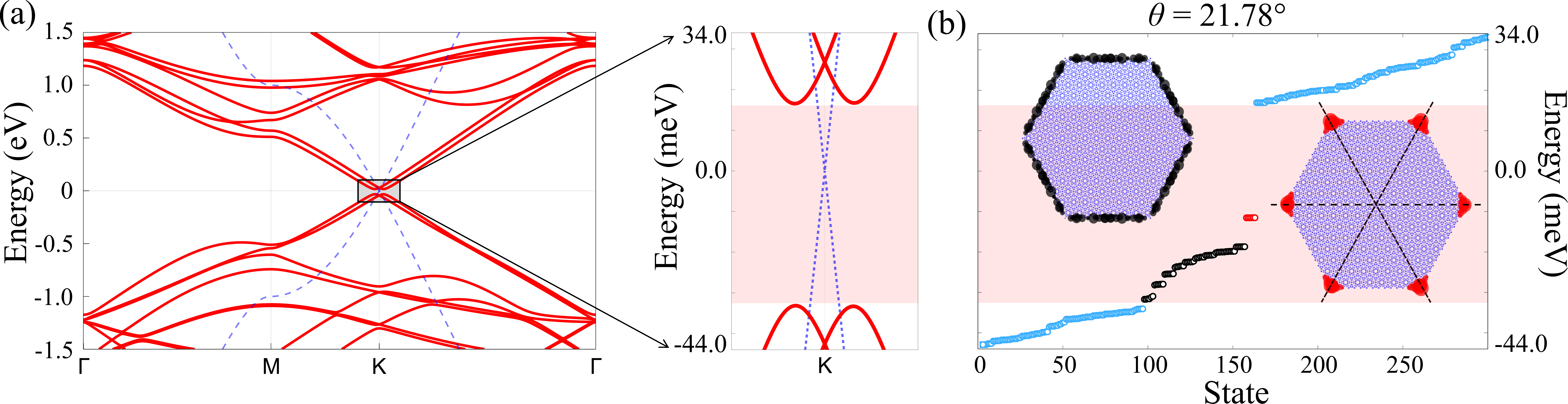}
    \caption{Electronic properties and higher-order topology at $\theta=\ang{21.78}$. (a) Band structures of the TBKL system at \ang{21.78}. The right panel shows the magnified view of the region near $K$ marked by a gray box in the left. The pink area represents the interval of the bulk band gap. The blue dotted line represents the bands of monolayer kagome lattice. (b) Discrete energy spectrum and the distributions of corner states for a hexagonal nanosheet. The corner and edge states are marked in red and black,
respectively. The spatial distribution of the corner and edge phonon modes are shown as the inserts. The dashed lines denotes mirror-invariant planes.
    \label{fig2}}
\end{figure*}
\textit{Lattice model and construction of Hamiltonian}.---
The monolayer kagome lattice can be spanned by two primitive vectors of a triangular Bravais lattice, i.e., $\bm{a}_{1,2} = a(\frac{\sqrt{3}}{2},\pm\frac{1}{2})$, where $a$ is the lattice constant~\cite{Elser1989,Kiesel2013}. 
Here, we consider a bilayer kagome lattice in the initial AA stacking with an interlayer distance $h$~\cite{CrastodeLima2019}, shown in Fig.~\ref{fig1}(a). 
By rotating the bottom layer in the plane relative to the top layer by an angle $\theta$  around the axis through the center of the hexagon of the bottom kagome lattice, we can construct a TBKL [see Fig.~\ref{fig1}(b)]. We found that when performing rotations, multiple axes can be chosen, and different choices of rotation axes result in different lattice structures and unit cells of TBKL. After analysis, we identified three rotation centers: hexagonal center, triangular center, and atomic lattice point center. These results are presented in Table~\ref{Tab1}.
Accordingly, the lattice vector $p\bm{a}_1$+$q\bm{a}_2$ of the rotated layer is changed into $q\bm{a}_1$+$p\bm{a}_2$. To ensure the periodicity of the TBKL, the adjustable rotation angle $\theta$ should satisfy
\begin{align}
    \theta &= \arccos\frac{p^{2}+q^{2}+4pq}{2\left(p^{2}+q^{2}+pq\right)}, \label{Eq.1}
\end{align}
where two coprime positive integers ($p$, $q$) determine the rotation angle $\theta$. The number of atoms $N$ in the commensurate unit cell in the TBKL system can be described by $N = 6(p^{2}+q^{2}+pq)$.

We now introduce a tight-binding Hamiltonian to describe the TBKL as
\begin{align}
    H_{\mathrm{TBKL}} = -\sum_{\braket{i,j}} t_{ij} c_i^\dagger c_j - \mu\sum_i  c_i^\dagger c_i- V \sum_{\substack{i\in \text{T} \\ j \in \text{B}}} (c_i^\dagger c_i - c_j^\dagger c_j) , \label{Eq.2}
\end{align}
where $c_i^\dagger$ ($c_j$) is the creation (annihilation) operator on site $i$ ($j$) for the two layers. The first term in Eq. (\ref{Eq.2}) represents the intralayer and interlayer electron hopping, and $\braket{i,j}$ runs over all the nearest-neighbor and next nearest-neighbor sites. The hopping amplitudes $t_{ij}$ read
\begin{equation}\label{Eq.3}
  t_{ij}=\left\{
             \begin{array}{ll}
             t\mathrm{e}^{-\beta(d_{ij}-a)}, &  \ \mathrm{for  \ the  \ intralayer \ intereaction},  \\
             t_z\mathrm{e}^{-\beta_z(d_{ij}-h)}, & \ \mathrm{for  \ the  \ interlayer \ intereaction},
             \end{array}
\right.
\end{equation}
where $d_{ij}$ denotes the distance between the sites $i$ and $j$,  $\beta$ ($\beta_{z}$) is decaying factor for the intralayer (interlayer) interaction \cite{Moon2012}. This approach allows us to accurately describe both intralayer and interlayer hoppings within the system. The second term in Eq.~(\ref{Eq.2}) takes into account the chemical potential $ \mu $, a universal modification to the on-site energy. The third term in Eq. (\ref{Eq.2}) is the interlayer potential difference $V$ between the the top (T) and bottom (B) layers. In the body content, we take $ t = 1 $ eV, $ t_z = 0.3t$, $\mu=-1$ eV, $\beta=20/a$,  $\beta_z=20/h$ and $V=0.012$ eV, which are fitting parameters determined through relevant experimental tests~\cite{CrastodeLima2019,Ye2018}. These parameters do not affect the topological properties of the system. Relevant parameter tests can be found in the Supplemental Material (SM)~\cite{SM}.

\textit{Band structures and higher-order band topology}.---In the absence of twist, the bilayer kagome lattice with AA stacking exhibits two Dirac cones exactly at corners of the original Brillouin zone, and another two Dirac cones near the Brillouin zone corners~\cite{CrastodeLima2019}. The latter two originate from the crossing of the Dirac-like bands from different kagome layers, so the twist induced intervalley scattering can gap out them once the twist occurs~\cite{Park2019, SM}. For TBKL systems under different twist angle $\theta$, we can obtain the electronic band structures and band gaps by diagonalizing the tight-binding Hamiltonian Eq.~(\ref{Eq.2}). We summarize the values of band gaps corresponding to different twist angles $\theta$ in Table \ref{Tab1}. The results show that the two Dirac cone near $K$ points open a visible band gap. The symmetry of TBKL systems with a rotation angle $\theta$ is equivalent to that of a twist angle ($\ang{60}-\theta$), so the maximum twist angle $\theta$ cannot exceed $\ang{30}$.
Moreover, it is found that the relatively large band gaps are related to the large-angle twist, i.e., \ang{21.78} and \ang{27.80}, which is in accordance with the result of TBG system \cite{Park2019}. The calculated band structure of the TBKL at $\theta=\ang{21.78}$ is shown in Fig. \ref{fig2}(a), which corresponds to a maximum band gap of 50 meV. In the following, we focus mainly on the nontrivial topological features of $\theta=\ang{21.78}$ in the main text.

\begin{table}
    \centering
    \renewcommand\arraystretch{1.2}
    \caption{The number of atoms ($N$) in the unit cell , the calculated energy gap and the higher-order band topology of the AA-stacking TBKL systems with different commensurate twist angles $\theta$. There are three types of centers of rotation here: the center of a hexagon, the center of a triangle, and the atomic lattice point, represented by $\hexagon$, $\triangle$, and $\bullet$ respectively. Here, the HOTIs are denoted by the notation "{\color{red}\checkmark}"}
    \begin{ruledtabular}
    \begin{tabular}{cccccc}
     ($p$, $q$) & $\theta$ & $N$ &  Center of rotation   & HOTIs & Gap (meV) \\
        \colrule
         $(1,2)$   & \ang{21.78}  & 42  & $\hexagon$   & \color{red}\checkmark  & 50.0       \\
                   &              &     & $\triangle$/$\bullet$ &  \ding{55}    & $\sim$    \\
         $(1,3)$   & \ang{27.80}  & 78  & $\hexagon$/$\bullet$  & \color{red}\checkmark  & 16.0    \\
                   &               &       & $\triangle$  &  \ding{55}          & $\sim$      \\
         $(2,3)$   & \ang{13.17}   & 114   & $\hexagon$   & \color{red}\checkmark  & 6.0        \\
                   &               &       & $\triangle$/$\bullet$   & \ding{55} & $\sim$      \\
         $(1,5)$   & \ang{17.89}   & 186   & $\hexagon$/$\bullet$   & \color{red}\checkmark  & 1.3   \\
                   &               &       & $\triangle$   & \ding{55}  & $\sim$     \\

         $(1,6)$   & \ang{15.18}   & 258   & $\hexagon$   & \color{red}\checkmark  & 0.3        \\
                   &               &       & $\triangle$/$\bullet$   & \ding{55}  & $\sim$        \\
    \end{tabular}
    \end{ruledtabular}
    \label{Tab1}
\end{table}

Next, we demonstrate that the HOTI phase appears in the TBKL system with the twist angle $\theta=\ang{21.78}$. We construct a hexagonal-shaped TBKL nanosheet with 380280 atoms. The calculated discrete energy spectrum is shown in Fig.~\ref{fig2}(b). For simplicity, we here only plot 300 states around the energy gap. Obviously, it exhibits visible edge states (marked by black dots) located in the energy gap, and the spatial distribution of edge states is illustrated at the edge of the TBKL nanosheet in the left inset of Fig. \ref{fig2}(b). More importantly, there are six degenerate in-gap corner states (red dots), which are distributed on six corners of the TBKL nanosheet, as shown in the right panel of Fig.~\ref{fig2}(b). The presence of corner states is a definitive characteristic of a 2D second-order topological insulator, protected by both the three-fold rotational symmetry ($C_3$) and mirror symmetry ($\sigma_{v}$).
The topological corner states in rhombic nanosheets with only the $\sigma_{v}$ symmetry and in triangular nanosheets with only the $C_3$ symmetry are presented in the SM~\cite{SM}.

The higher-order band topology can be characterized by the real Chern number $\nu_\text{R}$, i.e., the second Stiefel-Whitney number, which is defined by the parity eigenvalues of occupied bands at four time-reversal invariant momenta $\Gamma_i=(\Gamma, M_1,M_2,M_3)$ as \cite{Park2019,PhysRevLett.128.026405}
\begin{align}
(-1)^{\nu_\text{R}}=\prod_i(-1)^{[(n^{\text{odd}}_{\Gamma_i}/2)]}, \label{Eq.4}
\end{align}
where the $n^{\text{odd}}_{\Gamma_i}$ is number of occupied states with odd parity at each $\Gamma_i$. For the considered 84 occupied bands, there are 40 (38) bands with odd parity at $\Gamma$ ($M_1,M_2,M_3$), resulting in the parities at $(\Gamma, M_1,M_2,M_3)=(+,-,-,-)$. Therefore, the real Chern number $\nu_\text{R}=1$, which signifies the presence of bulk topology and confirms the topological origin of the corner states in the TBKL system.

Another physical manifestation of 2D SOTIs is the fractionally quantized corner charges \cite{Benalcazar2017,Benalcazar2017a,Ezawa2018,Peterson2018,PhysRevB.48.4442,Takahashi2021} protected by the crystalline rotation symmetry. 
Here, we employ the concept of rotation topological invariants $\begin{bmatrix}\Pi_p^{(n)}\end{bmatrix}$ (RTIs) to study the corner states filled by fractionally quantized corner charges.
The RTIs with $n$-fold rotational symmetry $C_n$ can be evaluated as \cite{Peterson2020,Song2017,Benalcazar2019,Schindler2019,Takahashi2021}
\begin{align}
 \begin{bmatrix}\Pi_p^{(n)}\end{bmatrix}\equiv\#\Pi_p^{(n)}-\#\Gamma_p^{(n)},\label{Eq.5}
\end{align}
where $\#\Pi_{p}^{(n)}$ ($\#\Gamma_p^{(n)}$) is the number of occupied states at the rotation-invariant high-symmetry points $\Pi$ ($\Gamma$) of the BZ, and $\Pi_p^{(n)}$ ($\Gamma_p^{(n)} $) represents the eigenvalues $\mathrm{e}^{2\pi\mathrm{i}(p-1)/n}$ of rotation operators $C_n$.
In the TBKL system, the combination of $C_3$ symmetry and $\sigma_v$ symmetry results in  the $C_6$ symmetry. For a $C_6$ invariant system, the symmetry forces equal representations at $M_1$, $M_2$ and $M_3$, as well as at $K_1$ and $K_2$, leading to two linearly independent RTIs, i.e., $[K_1^{(3)}]$ and $[M_1^{(2)}]$. In addition, the rotational center of the TBKL system is located at the 1$a$ Wyckoff position (the center of the $C_6$-symmetric unit cell). Therefore, the corner charge $Q_{1a}^{(6)}$ in a $C_6$ symmetric TBKL nanosheet can be represented by \cite{Takahashi2021}
\begin{align}
Q_{1a}^{(6)} &\equiv\frac{|e|}{6}\Big(n_{1a}^{(\mathrm{ion})}-v+2[K_1^{(3)}]+\frac{3}{2}[M_1^{(2)}]\Big)\pmod{e} \label{Eq.6},
\end{align}
where $n_{1a}^{(\mathrm{ion})}|e|$  and  $v$ represent the charge of the ion at the center position 1$a$ and the number of occupied bands, respectively. Based on Eqs. (\ref{Eq.5}) and (\ref{Eq.6}), we calculate the corner charge of the nanosheets. The obtained parameters $n_{1a}^{(\mathrm{ion})}=0$, $v=14$, $[K_1^{(3)}]=-2$, and $[M_1^{(2)}]=-2$ indicate that the value of corner charges distributing over six corners are $e$/2 (mod $e$), characteristic of the corner states fractionally filled. Note that we here only elaborate the AA-stacking TBKL system with twist angle $\ang{21.78}$. In fact, inspection of TBKL systems under other rotation angles listed in the Table~\ref{Tab1} reveals that they all exhibit the 2D HOTI behavior. Moreover, it is found other stacking ways, such as AB and AC stackings~\cite{CrastodeLima2019}, also lead to HOTI phases with corner states. This indicates that different stacking methods do not affect the emergence of higher-order band topology, further demonstrating the universality and generality of TBKL systems.

\textit{Effect of disorder}.---
To investigate the robustness of topological corner states, we next discuss the disorder effect on the TBKL system.  We apply a random potential term $H_\text{d}$ to each atom in a giant TBKL nanosheet, resulting in the Hamiltonian of TBKL system incorporated with disorder as
\begin{align}
    H_{\mathrm{TBKL}}^{\mathrm{dis}} = H_{\mathrm{TBKL}}+H_\text{d}= H_{\mathrm{TBKL}}+ \sum_i\xi_i c_i^{\dagger} c_i , \label{Eq.7}
\end{align}
where the random function $\xi$ is randomly distributed within an interval of $[-w,w]/2$, and $w$ is the disorder strength. Here, we take $w = 50$ meV, which is approximately in the same magnitude of the band gap of TBKL with the twist angle $\theta=\ang{21.78}$.

\begin{figure}
    \centering
    \includegraphics[scale=0.06]{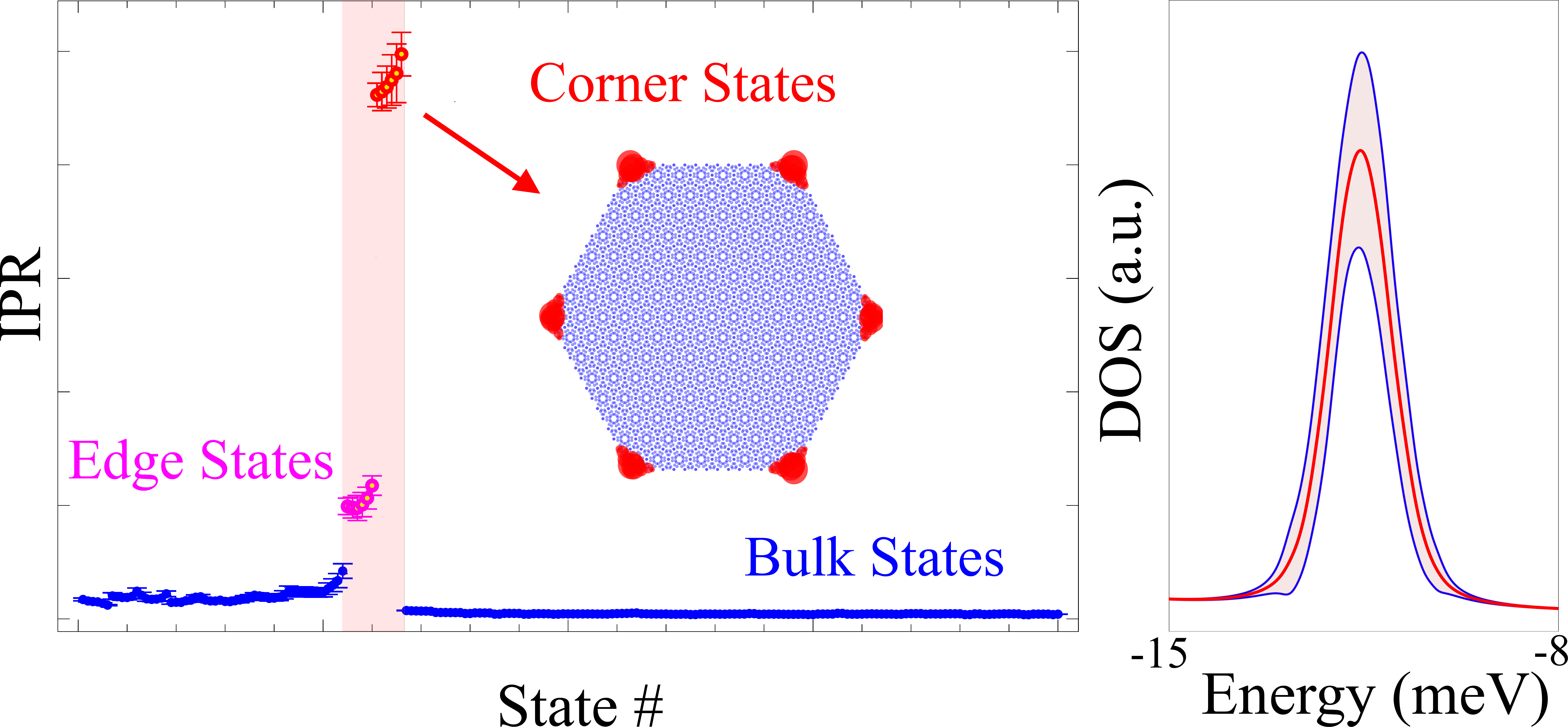}
    \caption{The electronic properties of the TBKL nanosheet with $C_6$ symmetry under the effect of disorder. (a) The IPR spectrum of the nanosheet under the disorder, in which 200 states near the Fermi level are plotted. We see that the degree of localization of the six corner states (red dot) is very high and stable in the presence of  disorder. It is worth noting that the deviation of the corner state is relatively small, indicating the robust higher-order corner states, while the edge states is relatively fragile due to the small IPR. (b) The DOS around the corner states, in which the blue lines represent the upper and lower limit of the DOS under the effect of disorder, and the red line denotes the average DOS. We find that the sixfold degenerate corner states are shifted by the disorder effect in DOS, but always located in the gap.
    \label{fig3}}
\end{figure}
To show the nature of each states, we employ the inverse participation ratio (IPR), which quantitatively characterizes the localization of the electronic states \cite{PhysRevB.54.10284,Konstantinou2019}.  
The calculated IPR spectrum of 200 electronic states near the Fermi level is shown in Fig. \ref{fig3}(a). It can be clearly seen that, with the large values of IPR, the six corner states are highly localized in the band gap [see the inset of Fig. \ref{fig3}(a)]. This suggests the robustness of the HOTI against disorder in the TBKL. We also calculate the density of states (DOS) around the corner to characterize the existence of corner states \cite{Liu2022z}. The DOS can be obtained via
\begin{align}
\mathrm{DOS}(\epsilon)=-\frac{1}{\pi}\text{Im}\sum_{i=1}^{n}\frac{1}{\epsilon-{E}_{i}+\text{i}\eta}\label{Eq.8},
\end{align}
where ${E}_{i}$ is the energy of the $i$th state, $n$ is the total number of states, and $\eta$ is an infinitesimal quantity. In the numerical calculation, we take $\eta$ = $0.01$ meV. Figure \ref{fig3}(b) shows the average DOS around the corner states. Although we introduce a sizable disorder with respect to the band gap, the presence of the typical peak, which corresponds to the corner charge in real space, illustrates that the electronic states are mainly located around the corner states, further implying the corner states of the TBKL system are inherently robust to disorder.

\textit{Effect of Charge Density Wave}.---As is well known, a $2\times2$ charge density wave order (CDW) that occurs in the kagome atomic layer
 was been demonstrated to be an important feature in kagome materials~\cite{Wilson2024,Lin2021,PhysRevLett.127.217601,Nie2022,FENG20211384,Tan2021,Yu2021,Kang2019,Xie2021}. The CDW order in kagome lattice is considered to have a significant influence on its electronic properties, such as the observed anomalous Hall effects. Therefore, it is desirable to uncover the interplay between the higher-order band topology and the CDW order in the TBKL system.

\begin{figure*}
    \centering
    \includegraphics[scale=0.087]{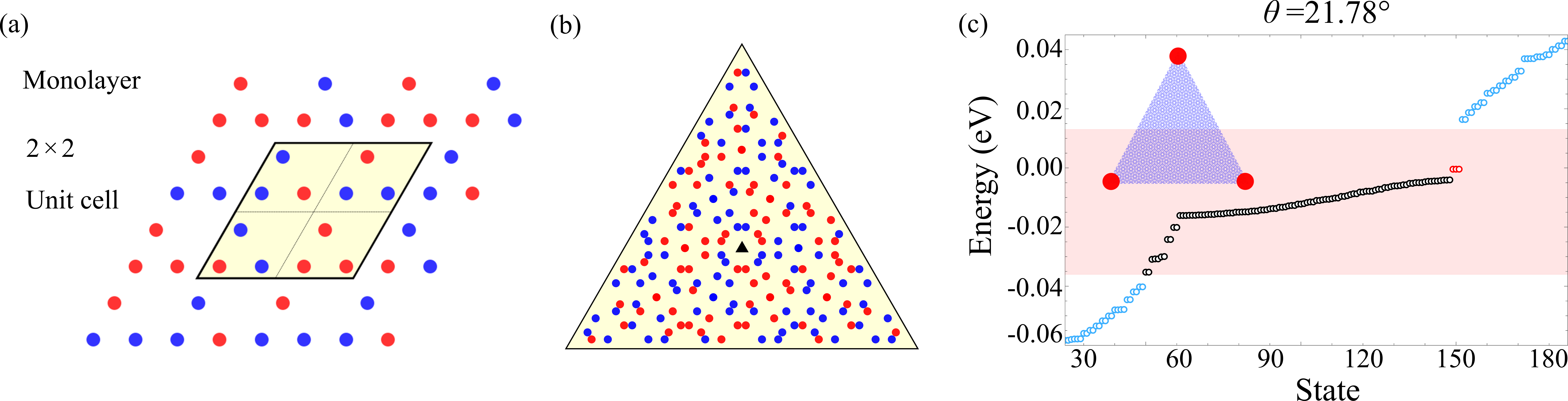}
    \caption{(a) The 2$\times$2 unit cell of the monolayer kagome lattice, and the black dashed lines represent these four basic unit cells. (b) A finite size triangular nanosheets of TBKL with $C_3$ symmetry under the effect of 2$\times$2 CDW order ($\theta=\ang{21.78}$).
    The blue and red atoms have  opposite on-site energy modification to the chemical potential ($-\mu \pm \lambda^\text{vCDW}$). The black triangle in the image represents the rotational axis of $C_{3z}$. (c) Discrete energy spectrum of a finite-sized triangular nanosheet with $C_3$ symmetry, in which three red dots represent the in-gap corner states, and the black ones denote the edge states.
    \label{fig4}}
\end{figure*}

Here,we mainly focus on the charge order as the vector charge density wave (vCDW), whose Hamiltonian reads $H^{\mathrm{vCDW}}= \sum_i \lambda_i^\mathrm{vCDW} c_i^{\dagger} c_i$, where $\lambda_i^\mathrm{vCDW}$ is the order parameter and describes the modification of on-site energy. Thus, the Hamiltonian of the TBKL system with vCDW order can be written as
\begin{align}
    H_{\mathrm{TBKL}}^{\mathrm{vCDW}} = H_{\mathrm{TBKL}}+ H^{\mathrm{vCDW}}. \label{Eq.9}
\end{align}
The inclusion of CDW order expands the unit cell of the monolayer kagome lattice, breaking the original $C_6$ symmetry down to $C_3$ symmetry [see Fig. \ref{fig4}(a)].  Additionally, the vCDW breaks the $\sigma_v$ symmetry. In the TBKL system in the absence of vCDW, we have shown that the $C_3$ symmetry and $\sigma_v$ symmetry can independently protect the higher-order topology as discussed above. Despite the breaking of $\sigma_v$ symmetry by vCDW, it is expected that the higher-order corner states related to the $C_3$ symmetry can be preserved. For related test calculations, please refer to the SM~\cite{SM}.
After considering the vCDW order, the moir\'e unit cell becomes larger and has the number of atoms $N=168 (42\times 4)$. In this case, the band gap retain the value of 50 meV [see Fig. \ref{fig4}(c)]. By constructing a triangular nanosheet with the $C_3$ symmetry, we can calculate the energy spectrum of the TBKL system with vCDW order, as shown in Fig. \ref{fig4}(c). The three degenerate states, which corresponds to the corner states (marked by red dots) located at the vertices of the triangular nanosheet [shown in the left inset of Fig. \ref{fig4}(c)], can be clearly seen in the band gap, confirming that the higher-order band topology of the TBKL system is robust against the CDW order.

\textit{Conclusion}.---In conclusion, by employing a tight-binding minimal model Hamiltonian, we theoretically predict the existence of  higher-order topological insulating phases in TBKL systems, which are independently protected by either the $C_3$ symmetry and $\sigma_v$ symmetry. Notably, we observe a substantial band gap of 50 meV in the TBKL system with a twist angle of \ang{21.78}. The higher-order topology is characterized by the real Chern number $\nu_\text{R}=1$ and the fractionally quantized corner charges are $e$/2 (mod $e$). Importantly, our findings demonstrate the robustness of the higher-order topological corner states even in the presence of disorder and the 2 $\times$ 2 vCDW order. It is worth noting that the band gap that opens up near $K$ point is closely linked to the readily adjustable twist angle, providing an effective means to control the higher-order band topology and potentially realize HOTIs in experimental settings. We anticipate that our work will contribute to the advancement of readily controllable HOTIs in 2D twisted systems.

\bibliographystyle{apsrev4-1}
\bibliography{ref}

\end{document}